\documentclass[twocolumn,amsmath,showpacs,amssymb,aps,nofootinbib,floatfix]{revtex4}
\usepackage{epsfig,amsmath,amssymb}
\usepackage{dcolumn,color}
\usepackage{bm}
\usepackage{graphicx}

\newcommand{\beq}{\begin{equation}}
\newcommand{\eeq}{\end{equation}}
\newcommand{\eqcomma}{\phantom{A},\phantom{A}}
\newcommand{\order}[1]{ \mathcal{O} \left( #1 \right) }
\newcommand{\ave}[1]{\left\langle #1 \right\rangle}

\begin{document}

\title{The ideal relativistic fluid limit for a medium with polarization}
\author{David Montenegro$^1$, Leonardo Tinti$^2$, Giorgio Torrieri$^1$}
\affiliation{$^1$ IFGW,Unicamp, Campinas, Brasil}
\affiliation{$^2$Department of Physics, The Ohio State University, Columbus, OH 43210-1117, USA}
\date{\today}

\begin{abstract}
We use Lagrangian effective field theory techniques to construct the equations of motion for an ideal relativistic fluid whose constituent degrees of freedom have microscopic polarization.   We discuss the meaning of such a system, and argue that it is the first term in the EFT appropriate for describing polarization observables in heavy ion collisions, such as final state particle polarization and chiral magnetic and vortaic effects.  We show that this system will generally require non-dissipative dynamics at higher order in gradient than second order, leading to potential stability issues known with such systems.  We comment on the significance of this in the light of conjectured lower limits on viscosity. 
\end{abstract}

\maketitle
\section{Introduction}
Relativistic hydrodynamics is a topic of very active theoretical and phenomenological 
development \cite{kodama}.  Phenomenologically, it seems to provide a good description of physics in heavy
ion collisions, making numerical hydrodynamic solvers an indispensable tool in this field.  
One phenomenon which has not been taken into account dynamically in hydrodynamics is polarization. This is understandable because it is not immediately straight-forward to link it to flow observables. Hydrodynamics deals with macroscopic, coarse grained quantities, while particle polarization is a microscopic one. More over that spin polarization is a particle physics concept while hydrodynamics can be defined (and it is used) even when there are no known (quasi-) particle excitations. 
The direct observation of polarization of $\Lambda$ particles in heavy ion collisions \cite{upsal} has  the potential to change this. A reliable set of tools to analyze this observable theoretically is however still lacking.  For instance Israel-Stewart equations (the most commonly used for second order viscous hydrodynamics) is usually thought of as a limit of the relativistic Boltzmann equation. Spin dynamics has been completely neglected in the derivation, appearing only as a degeneracy factor, which is equivalent to an assumption of equiprobability of polarization. This is at odds with statistical equilibrium with non-vanishing angular momentum \cite{becattini1,becattini2,becattini3}.  

Polarization, vorticity and chirality observables have received some amount of attention in literature in the context of vorticity-induced polarization \cite{uspol,wangpol,csernaipol}, hadronic reactions \cite{rafpol,hamapol} and generic transport theory including chirality \cite{stephanov1,stephanov2,igor,shocks}, the latter motivated by the hypothesis of the chiral magnetic effect \cite{magnetic1} and its hydrodynamic  \cite{surowka} and magnetohydrodynamic \cite{kovtunm,saso} incarnation.   Note that spin-orbit coupling, unlike the effects described here, is not anomalous, since vortical susceptibility is directly related to spin-orbit coupling \cite{vortsus}.   That said, if the system arising in heavy ion collisions is an ideal fluid and polarization is non-negligible, anomalous transport, as a deviation from local equilibrium, should be sub-leading to the effect of spin-orbit interactions in an evolving fluid.

An effective theory for describing the relationship between vorticity and polarization is however still missing.   The problem is that vorticity does not emerge in the transport limit, but rather close to the thermodynamic and hydrodynamic regime.   Thermodynamics was studied within the usual  techniques, updated with the inclusion of angular momentum \cite{becattini1,becattini2,becattini3}, but this is not generally a good approximation for a strongly coupled dynamical system, where equilibrium is local rather than global.  \cite{csernaipol} used thermodynamic equilibrium within an isochronous Cooper-Frye formula assuming polarization is zero before freezeout, but this assumption violates detailed balance across the freezeout hypersurface \cite{csernaifreezeout}.   If hadrons after freeze-out carry both vorticity and spin polarization, then so must the constituents of a fluid before freezeout.   Such a fluid, however, while being studied in condensed matter systems \cite{spintronics,spintronicstheo} still needs to be developed for the ultra-relativistic limit.   Even intuitively, the idea that only quasi-particles carry polarization makes envisioning such a fluid confusing.

The problem is a conceptual one: local {\em isotropy}, one definition of an ideal fluid, {\em forbids} the transfer of angular momentum to spin as that would create an anisotropy independent of the coarse-graining scale \cite{uspol} (such anisotropy, in normal hydrodynamics, is directly proportional to the mean free path).   However, local {\em equilibration} and entropy maximization in the presence of angular momentum and spin explicitly necessitates of such an anisotropy \cite{becattini1}.  And, of course, vorticity conservation, the Noether current of the diffeomorphism invariance underlying perfect fluid dynamics \cite{nicolis2}, should be broken if angular momentum can be transferred to local polarization.   While attemps to resolve this issue go back decades \cite{oldspin1,oldspin2}, the contradiction between the various definitions of hydrodynamics can still elicit confusion.

To clarify this situation, one can turn to microscopic transport theory \cite{huang}.   Polarization is {\em not} a purely transport phenomenon since it can occur in equilibrium if the system aquires angular momentum.
However, non-equilibrium Microscopic polarization is a violation of molecular chaos, since it means that the distribution function is generalized into polarization components $f(x,p) \rightarrow \left\{ f_i(x,p)  \right\}$ whose correlation cannot be factorized, $\ave{f_i f_j} \ne \ave{f_i}\ave{f_j}$ within a microscopic volume element.  It is describeable fully within only higher terms of the BBGKY hyerarchy w.r.t. the average.
In the strongly coupled limit, boson-fermion couplings (the interaction between color $D3$ and higher order flavor branes \cite{d3d7}) are suppressed by factors of $N_c$ (following the hyerachy studied in \cite{witten}), leading support to the idea that this qualitative picture also applies at strong coupling (transverse polarization of the spin degrees of freedom in $N=4$ SYM is a more subtle issue since gauge invariance there has a non-trivial effect.  To calculate this exactly one would need to calculate the 2-point function of vorticity beyond leading order, a work in progress).

To summarize this argument, as is well known hydrodynamics is based on a hierarchy of three length scales \cite{ll,betzfluid,torrieri1,torrieri2,torrieri3,glorioso}
\begin{equation}
\label{scales}
n^{-1/3} \ll l_{mfp} \ll \left(\partial u_\mu \right)^{-1}
\end{equation}
where $n^{1/3} \equiv T_0$ is the separation of the microscopic degrees of freedom (whose effect are non dissipative but probabilistic \cite{torrieri1}), $l_{mfp}$ is the mean free path (whose effect is deterministic but dissipative) , and $\partial u_\mu$ is the gradient of the velocity.  The second inequality controls leading-order dissipative phenomena, such as viscosity, conductivity and sound attenuation.  The first inequality regulates the departure from molecular chaos, the irrelevance of thermodynamic fluctuations to hydrodynamic evolution.  

If one considers hydrodynamics as a gradient expansion, as is done in \cite{nicolis3}, one generally forgets the first inequality and uses the second inequality to define an effective theory expansion parameter (the Knudsen number) which can be developed into a gradient effective theory. 
The current of the angular momentum with respect to the point $x_0$ reads
\begin{equation}
\label{angcurr}
 M^{ \lambda \mu  \nu}_{x_0} =  \left[ (x-x_0)^\mu T^{\lambda \nu} - (x-x_0)^\nu T^{ \lambda \mu} \right].
\end{equation}
In which $T^{\mu\nu}$ is the expectation value of the stress-energy tensor (before coarse-graining), defined as the derivative of the action with respect to the metric tensor. At global equilibrium, the flux of this, namely the total angular momentum, is given by the total four momentum and the vorticity.
In the ideal hydrodynamic limit the latter is linked to the circulation, which is used to define, trough the Stokes theorem, another antisymmetric rank two tensor also called vorticity.  This is a conserved quantity in the usual perfect fluid approach, whose symmetry is the theory's volume preserving diffeomorphism invariance \cite{endlich,nicolis2}. In order to avoid confusion we will refer to this definition of vorticity as circulation, even if we refer to the rank two tensor and not to the line integral.
If we insert polarization effects in ideal hydrodynamics, one can expect that (as we will show later) the new degrees of freedom could act as a source of circulation, therefore breaking explicitly the circulation theorem. This can be considered as a feed back of the polarization degrees of freedom (order zero in a gradient expansion) to the circulation (linked to the geometric vorticity, even if it the two are not exactly proportional in general) which is expressed as gradients of the hydrodynamical variables.  Hence, one would expect the gradient expansion to break down in a very specific way, which should be related to the enriched structure of the conserved angular momentum, the divergence of Eq. ~\ref{angcurr}, if polarization effects can't be neglected.
Unlike a normal EFT expansion,  however, the appropriate terms will not be generic higher order gradient terms, but will be precisely constrained by symmetries (as we show in \cite{prl}, this can be clearly seen in the linearized limit when the dispersion relation is considered), and will coincide with the gradient terms expected in the global equilibrium state with angular momentum \cite{becattini2}.   

To quantify the latter, 
in a relativistic setting with a small chemical potential, our intuition tells us that the microscopic density of degrees of freedom  $(gT)^{3}$ where $g\sim N_c^2$ is the microscopic degeneracy (numerically $(gT)^{3}\times \mathrm{Volume}$ is $\order{10^{2-3}}$ in heavy ion collisions, $\order{10^4}$ in ultracold atom systems) might increase the amount of angular momentum stored microscopically by equipartition.   However, experience with magnets tells us that in the high temperature limit the net polarization decreases as $\sim \tanh(\mu_S/T)$ where $\mu_S$ is the microscopic polarizeability and, generically, spin-orbit couplings between gauge bosons and Fermions in the fundamental representation are suppressed in the planar limit.

Summarizing these considerations, one naively expects polarization in a hydrodynamic system to be small (\cite{xie} argues parametrically smaller than the total vorticity), but not necessarily parametrically smaller than the mean free path (the experimental measurement of polarization in a system commonly thought to be hydrodynamic confirms this expectation).    We therefore aim to see how the gradient expansion is altered in the limit when polarization is non-negligible.  As we show in the next section, combining the EFT picture with the symmetry properties of angular momentum can accomplish this.
\section{Hydrodynamics as an effective theory}
The theoretical tools necessary to develop hydrodynamics in this limit \cite{nicolis1,torrieri1,torrieri2,torrieri3,glorioso}, and to relate it to dissipation \cite{grozdanov} are well known: One writes down hydrodynamics in lagrangian form, and treats the microscopic scale as an effective Planck constant \cite{torrieri1}, the Knudsen number as an effective theory scale hyerarchy and develops fluctuation-driven terms within the effective field theory.  

A perfect fluid without polarization can be described by three fields $\phi^I$, describing the three Lagrangian coordinates of the systems.
The fact that it is a fluid  can be imposed through a volume-preserving
diffeomorphism invariance  \cite{nicolis1,nicolis2,nicolis3}
\begin{equation}
\label{diffeoinv}
L(\phi_I \rightarrow \xi_I(\phi_J)) \rightarrow L \eqcomma 
 \det\left[ \frac{\partial \xi_I}{\partial \phi_J }\right]=1
\end{equation}
  It therefore follows that the Lagrangian with the lowest order possible of gradients is of the form (notation of \cite{nicolis3}) 
\begin{equation}
L=F(b) \eqcomma b=\sqrt{\det_{IJ} \left[B_{IJ}\right]}\eqcomma B_{IJ}=\partial_\mu \phi_I \partial^\mu \phi_J
\end{equation}
The Lagrangian above can be shown in a straight-forward way to yield the energy momentum tensor whose conservation yields Euler's equations \cite{nicolis1,nicolis2,nicolis3},
\begin{equation}
  \label{en_mom}
\partial_\mu T^{\mu \nu}=0 \eqcomma T^{\mu \nu} = (p + e) u^\mu u^\nu - pg^{\mu \nu}.
\end{equation}
If one wants to include a chemical potential $\mu$, the energy density and pressure read instead \cite{nicolis3}
\begin{equation}\label{Fbmu}
e= \mu \frac{dF(b,\mu)}{d\mu} -F(b,\mu) \eqcomma p= F(b,\mu)-\frac{dF(b,\mu)}{db} b .
\end{equation}
Note that the lagrangian $F(b,\mu=0)$ coincides with the energy density for vanishing chemical potential and corresponds, in general, to a Legendre-transformed energy.
The chemical potential is also related to the Noether current generating the scalar conserved charge, a $U(1)$ symmetry, by
\begin{equation}
\label{noether}
L( \exp[i\psi]) \rightarrow L(  \exp[i(\psi + c)]) \eqcomma \mu=u_\mu \partial^\mu \psi.
\end{equation}
The flow velocity $u^\mu$ is defined as $u^\mu \partial_\mu \phi_J=0 \forall J$, which in four dimensions leads uniquely to a 4-vector normalized to unity
\begin{equation}
\label{uphidef}
u_\mu = \frac{1}{6 b} \epsilon_{IJK} \epsilon_{\mu \alpha \beta \gamma} \partial^\alpha \phi^I \partial^\beta \phi^J  \partial^\gamma \phi^K 
\end{equation}
with the comoving projector being
\begin{equation}
\label{projector}
\Delta^{\mu \nu} = g^{\mu\nu} - u_\mu u^\nu = B_{IJ}^{-1} \partial^\mu \phi_I \partial^\nu \phi_J ,
\end{equation}
where we used the mostly plus convetion for the metric tensor $g^{\mu\nu}$.
Since the four-velocity defined in~(\ref{uphidef}) is by construction the direction of a local conserved four-current, 
\begin{equation}
\label{entropy_current}
K_\mu = \frac{1}{6 } \epsilon_{IJK} \epsilon_{\mu \alpha \beta \gamma} \partial^\alpha \phi^I \partial^\beta \phi^J  \partial^\gamma \phi^K = b u_\mu  \Rightarrow \partial_\mu K^\mu = 0. 
\end{equation}
it is natural to identify it with the entropy current, since entropy is the only locally conserved current in a perfect fluid with no conserved charges.

The relativistic extension of the Kelvin circulation theorem, usually referred to as vorticity conservation, arises in this description as a non-local Noether current of the diffeomorphism invariance of the theory, specifically \cite{nicolis2,endlich}
\begin{equation}
\label{current}
 \oint_\Omega dx_i u^i \frac{dF(b)}{db} =- \int_0^1 d\tau  \times 
\end{equation}
\[\ 
\times \int d^3 x \frac{\partial L}{\partial (\partial_0\phi^I)} \frac{d\Omega^I}{d\tau} \delta^3 \left( \phi^J - \Omega^J(\tau) \right).
\]
That is the circulation of the three-velocity (times a function) along the flux tubes \footnote{The circulation is on the flux lines since $\Omega^I$ has components in the internal indices space, the Lagrange coordinates, and not the space coordinates themselves.} defined by the loop $\Omega^I$.
The LHS of the equation is one definition of vorticity, and the RHS is the Noether current
$\frac{\partial F}{\partial (\partial_\mu \phi^I)} \zeta^I_\Omega(\phi_I)$ for a, infinitesimal volume preserving diffeomorphism with generator\footnote{It can be proven that $\zeta^J_{\Omega}$ fulfills $\sum_I \partial\zeta^I/\partial\phi^I$ as it has to for being a generator of volume preserving diffeomorphisms.}
\begin{equation}
\label{zetadef}
 \zeta^I_\Omega(\phi^J) = - \int_0^1 d\tau \frac{d\Omega^I}{d\tau} \delta^3\left(\phi^J -\Omega^J(\tau) \right) 
\end{equation}
which moves coordinates among the loop between the parameter values $\tau=0$ and $\tau=1$.
\section{Polarization degrees of freedom}
If the system has intrinsic polarization, a net spin direction where a fraction of microscopic degrees of freedom points to, 
the coordinates $\phi^I(x)$ are not enough because they do not contain information about polarization.   

To find the appropriate additional degrees of freedom, we need to understand how to generalize hydrodynamics in a situation where some of what are considered fundamental principles of it, such as local isotropy, are inappropriate.
The principles we choose to use are
\begin{description}
\item[(i)] The dynamics within each cell is faster than macroscopic dynamics, and it is expressible only in term of local variables and with no explicit reference to four-velocity $u^\mu$ (gradients of flow are however permissible, in fact required to describe local vorticity).
  \item[(ii)] This dynamics is dictated by local entropy maximization, within each cell, subject to constraints of that cell alone.  In the ideal limit, macroscopic quantities are assumed to be in local equilibrium inside each macroscopic cell (even if gradients are not vanishing and the system can be relatively far away from global equilibrium).   This point is what distinguishes our approach from previous treatments, including the widely cited works in this subject from decades ago \cite{oldspin1,oldspin2}
\item[(iii)] The only excitations allowed around a hydrostatic medium are sound waves and vortices
\end{description}
We shall examine the consequences of each assumption in detail throughout the paper. 
The intrinsic angular momentum of a fluid cell is the integral around a small hypervolume $\delta\Sigma$ of the flux of angular momentum 
\begin{equation}\label{angmom}
 \delta J^{\mu\nu} (x) = \int _{\delta\Sigma} d\Sigma_\lambda M^{\lambda\mu\nu}_x
\end{equation}
According to principle (i), the integral is performed in the local comoving frame $d\Sigma_{\mu} = dV u_\mu$. Since angular momentum can be exchanged (in macroscopic time scales) between cells, this is not a conserved quantity. In order to have a polarization which is not infinitesimally small, we normalize to the small volume of the coarse graining $\delta V = \int_{\delta\Sigma}d\Sigma$

\begin{equation}
\Psi^{\mu\nu} (x) = \frac{\delta J^{\mu\nu}}{\delta V}.
\end{equation}
In this way the variable $\Psi^{\mu\nu}$ can be considered the local "angular momentum" of the fluid cell, which is usually neglected in a coarse graining procedure. As a mathematical simplification, during the remainder of this work we will assume that only the part of $\Psi^{\mu\nu}$ orthogonal to the four velocity will be the relevant one to be included in the effective Lagrangian treatment. This is the part related to the classical part of the angular momentum (the one which ends up in the generator of rotations, as opposed to the time-like one which is related to the boost generator), note how in Ref.~\cite{becattini1} this is the part which is actually responsible of particle polarization in global equilibrium and the weak coupling limit. We will call this variable $y^{\mu\nu}$

\begin{equation}
 y^{\mu\nu} = \Delta^\mu_\alpha \Delta^\nu_\beta \Psi^{\alpha\beta},
\end{equation}
where $\Delta^{\mu\nu} = g^{\mu\nu} - u^\mu u^\nu$ is the projector on the hyper-plane orthogonal to the local four-velocity $u^\mu$. 

It must be noted that $y^{\mu\nu}$ depends on the coarse graining scale. This is unavoidable because of the non-extensivity of the  angular momentum. However, contrary to what one might expect, it is not vanishing in the vanishing volume limit in the case of constituents with spin. Using the physical intuition from classical mechanics one expects for a system at global equilibrium an orbital part (which vanishes since moments of inertia over volume vanish in the small volume limit) and a polarization contribution.  This contribution was calculated explicitly in ref. \cite{becattini2} as seen from the lab frame, for a rotating gas of particles with spin in global equilibrium.   It shows a "spin component" of the angular momentum density that becomes constant in the non-relativistic limit. The "spin component" of the total angular momentum is almost proportional (exactly in the non-relativistic limit) to the volume, and in particular the ratio with the volume is not vanishing in the small volume limit. This provides an example of a physical situation where $y^{\mu\nu}$ is not vanishing.   The Lagrangian approach will allow us to study how $y_{\mu \nu}$ behaves when equilibrium is local rather than global.

\section{The effective theory Lagrangian}

According to point (ii) the fluid cells are suppsed to be in local equilibrium. It is known that in the case of thermodynamical equilibrium, angular momentum is proportional to the antisymmetric part of the four-velocity gradients~\cite{becattini1, becattini2}
\begin{equation}
  \label{vorttensor}
 \frac{1}{2}\left[  \partial_\mu u_\nu -\partial_\nu u_\mu\right] = \frac{1}{2}\left[   A_\mu u_\nu - A_\nu u_\mu \right] + \omega_{\mu\nu}, 
\end{equation}
being $A^\mu = \dot u^\mu = u^\rho\partial_\rho u^\mu$ the four-acceleration and $\omega_{\mu\nu}$ the vorticity (the space part of the antisymmetric part of the gradient).
Equilibrium, local or global, implies that the space part of 
polarization has to be in a one to one correspondence to the vorticity subject to entropy maximization.   More specifically, if sound waves and vortices are the only excitations within the hydrostatic limit this means polarization and vorticity have to point in the same direction, therefore
\begin{equation}
\label{prevortspin}
y_{\mu \nu} = \chi(b,\omega^2) \omega_{\mu \nu}
\end{equation}
This is a very general point:
 If we allow polarization and vorticity at equilibrium to be aligned by an angle $\theta \ne 0$, it would generate a broken continuus symmetry (the longitudinal angle $\varphi$ where the polar $\theta$ is defined by vorticity, could take any value), with Eq. \ref{prevortspin} updated to $y^{\mu \nu}=\chi(b,\omega^2) \Lambda^\mu_\alpha(\theta,\varphi) \Lambda^{\nu}_\beta (\theta,\varphi) \omega^{\alpha \beta}$ and $\Lambda^\mu_\nu (\theta,\varphi)$ is, in the comoving frame, a rotation matrix.  This carries with it a Goldstone mode (excitation in $\varphi$) with non-trivial topological structures (rings in $\varphi$ have to be continuos).   A detailed investigation of this model is left for another work, though we suspect that dissipative corrections, where polarization relaxes to vorticity, make the ideal limit of such a system un-realizeable.
 Beyond the ideal fluid limit, as discussed in \cite{denicolvort}, an Israel-Stewart approach will be likely where $y_{\mu \nu}$ {\em relaxes} to the vorticity tensor.

 We note that $y_{\mu \nu}$ is an auxiliary field interacting with $\phi_I$ rather than an extension of $\phi_I$ to internal degrees of freedom.
This reflects the fact that  spin is not conserved separately to angular momentum.   In Noether's language,  diffeomorphisms such as  Eq. \ref{diffeoinv} act on $\phi_I$ but not $y^{\mu \nu}$, since the latter represents an internal symmetry.    However, if we combine a Lagrangian invariant under Eq. \ref{diffeoinv} with a locally invariant term for the internal rotation diffeomorphisms, a conserved current comprising space and internal symmetries, physically represented by a current combining spin current and vorticity-carried angular momentum, would arise.
This can be seen more explicitly by considering that Eq. \ref{zetadef} only moves around $\phi_I$ and not $y^{\mu \nu}$ breaks the symmetry. For  instance, consider an effective Lagrangian of the form $F(b, \cdots y \cdots)$, where $\cdots y\cdots$ stands for the dependency of the Lagrangian density on a scalar combination of $y_{\mu\nu}$, its derivatives, and eventual mixed terms with the gradients $\partial_\mu\phi^I$ which respects all the previous symmetries. Hence, the conserved current corresponding to the loop $\Omega_J$ reads

\begin{equation}
 J^\mu_{\Omega} = \frac{\partial F}{\partial{(\partial_\mu\phi^I)}}\zeta^I + \mbox{higher order derivatives terms},
\end{equation}
the term proportional to $\partial F/\partial b$ will give one term which is again a circulation of a function times $u^i$, but the additional terms will add, in general, a circulation of an object which is not proportional to the velocity, preventing a further extension of the circulation theorem. For instance, looking in Sec.~\ref{dynamics} one can find one instance of the derivatives of the effective Lagrangian, and it is straightforward to check that in this case there is a, rather complex, source term for the circulation of $u^i$. In the end this means that the conserved current for the volume preserving diffeomorphisms can not be related to a familiar concept like the relativistic version of the circulation theorem.

In order to proceed one has to insert the new variables tracking microscopic polarization in the effecting Lagrangian. The combination has to be a scalar and, as a first attempt, we assume that only the lowest order in gradients will be needed. Hence, by counting gradients and enforcing symmetries, the lowest order scalar term is $y_{\mu \nu} y^{\mu \nu}$. For example $det[y]$ is a higher term in gradients, since $y^{\mu\nu}$ itself is proportional to a gradient of a macroscopic quantity by definition, $\epsilon_{\alpha \beta \gamma\rho} \partial^\mu K^\nu y^{\gamma \rho}$ would violate parity and $\partial_\mu K_\nu y^{\mu \nu}$ is proportional to $y_{\mu\nu}y^{\mu\nu}$.  Parity violating terms would of course be permitted in the context of anomalous hydrodynamics, but we will not consider it in the present work.  In order to handle more easily  the resulting equations we make use now of  some phenomenological ansatz, which can be however easilly relaxed if one wants to study the more general case. Considering that polarization introduces a correlation between microstates, the presence of polarization at a given entropy $b$ should change the free energy, to leading order in gradient, as $ b \rightarrow b \left(1-c y_{\mu \nu} y^{\mu \nu}  \right)$
where $c$ is a dimensionful constant representing polarizeability (it can be positive for a ferromagnetic material and negative for an antiferromagnetic one).   For dimensional reasons, and because of Eq \ref{scales}, $c \sim T_0^{2}$
Given this, a physically reasonable way to introduce polarization is
\begin{equation}
  \label{rescaledf}
  F(b,y) \rightarrow F\left( b\times f(y)  \right) \eqcomma f(y) = 1-c y_{\mu \nu} y^{\mu \nu} + \order{y^4},
  \end{equation}
where $y$ is a short-hand notation for $y_{\mu\nu}y^{\mu\nu}$.

In principle one should know the exact form of the effective Lagrangian in order to solve the equations of motion. However only a few (constant) parameters are necessary for the study of the small perturbations over a static background (Linearized theory). The next chapter will be dedicated to that. We will end this section explaining the possible ways to fix (without phenomenological assumptions like the last one, if needed) the form $F$, linking the Lagrangian formulation and usual thermodynamics using the methods of \cite{nicolis3} but with the angular momentum in lieu of chemical potential (note that the collinearity between angular momentum and polarization is what makes this analogy possible).

One has to be careful with this because, as illustrated for example in \cite{becattini2,Becattini:2014yxa}, entropy in systems with angular momentum is generally non-extensive.  Polarization is however intensive, as any other property dealing with microscopic properties.   Because of this, one cannot, as is usual in ideal hydrodynamics without polarization, 
assume the thermodynamic limit for the equation of state. However, one can still define local equilibrium within a microscopic cell in its comoving frame, starting from a finite-size statistical treatment \cite{becattini2,Becattini:2014yxa}.

By analogy with ref.~\cite{nicolis3} where it was said about the effective Lagrangian density $F(b,\mu)$ in Eq.~(\ref{Fbmu}) "It can be thought as a somewhat unusual thermodynamic potential where: $dF = -Tds +nd\mu$". We can expect that the Lagrangian density in Eq.~(\ref{rescaledf}) would correspond to thermodynamic potential 
\begin{eqnarray}
  \label{gd2}
d F (b,y) &=& \partial_b F \, ds + \partial_{y_{\mu\nu}}F \, d y^{\mu\nu} = \\ \nonumber
&=&  -(1-c y^2) F^\prime ds -2cb F^\prime y^{\mu\nu}dy_{\mu\nu}.
\end{eqnarray}
Note that this makes explicit the fact, inferred from (ii), that $y_{\mu \nu}$, while being a source of a conserved quantity, is not a dynamical degree of freedom, since the amount of angular momentum is not determined by initial conditions but rather by entropy maximization.     A solution with a boundary condition with different $y,b$ and velocity should generate the sort of shock-wave studied in \cite{shocks}.

In order to recognize the derivatives of this thermodynamic potential, one should do like in Ref.~\cite{nicolis3}. Namely compute the stress-energy tensor, consider $b=s$, since there is no other conserved vector current except the entropy density, and check which one is the form of $F$ that will reproduce the thermodynamical relations obtained from another source (for instance lattice gauge results for the equation of state of QCD).

In hydrodynamics without polarization, for instance, the analogous of Eq.~(\ref{gd2}) gives rise to the Gibbs-Duhem relation relating pressure and energy density to temperature and entropy density.   Here, because of the presence of angular momentum, terms like pressure and energy-density will not have such a simple relationship to actual energy and momentum flow within the fluid. Unfortunately it is not known the exact form of the entropy density for a spinning system, mainly because of the difficulty of computing the logarithm of the partition function for a generic relativistic-quantum system. In Ref.~\cite{becattini1, becattini2} in any case we can see that, in the weak coupling limit and for small vorticity, the angular momentum of a spinning system is proportional to the vorticity itself as in~(\ref{prevortspin}), an we can guess that

\begin{equation}
 -2cb F^\prime y^{\mu\nu} = -2cb \chi F^\prime \omega^{\mu\nu} \propto \frac{1}{T}
\end{equation}
being $T$ the local temperature.

The derivative of $F$ w.r.t. $y_{\mu \nu}$ is related to the vortical susceptibility in the way described in \cite{vortsus}.   As such, it will be strictly related to the magnetic susceptibility \cite{kovtunm,saso} (in one case one deforms $A_\mu$, in the other the perpendicular components of the metric \cite{vortsus}), and can be inferred from lattice results at finite magnetic field \cite{lattice}.  It can also be computed explicitly \cite{rotlat}.

  One important point to note is that we inserted the polarization related degrees of freedom in the effective Lagrangian and, in order to enforce local equilibration, we substitute $y^{\mu\nu}$ with a functional of the old degrees of freedom to enforce local equilibrium on each fluid cell. Thanks to~(\ref{prevortspin}) the effective Lagrangian becomes second order in the derivatives of the fundamental effective fields $\phi_I$. The reason to consider these gradients and not, for instance the symmetric part of the four-velocity gradient, lies in the assumption of local equilibration, namely point (ii). Contrary to most expectations, global equilibrium doesn't imply vanishing gradients. It does only in the case of homogeneous equilibrium (translationally and rotationally symmetric), however an average angular momentum is breaking rotation and translation invariance. In general equilibrium requires a timelike direction fulfilling a killing equation, see for instance Refs.~\cite{becattini1,kovtunm}. The case we are interested in is the one with average angular momentum, in this case it is straightforward to prove that the four-velocity may have only an antisymmetric gradient. This one, remaining at equilibrium, shouldn't be considered a dissipation inducing gradient, and therefore it can enter the effective Lagrangian for ideal hydrodynamics with polarization. The symmetric part however, being vanishing at equilibrium, should be safely considered a dissipation inducing term, like it has always been done in hdrodynamics, and it is reasonable to wait to extend the model to non ideal hydrodynamics with polarization before including it in the effective Lagrangian, suppressed by factors of the order of the Knudsen number.

  The equilibrium calculation with angular momentum can be used to justify the choice of considering only the space like part of the fluid cell internal angular momentum in the definition of $y^{\mu\nu}$, $i.e.$ the one proportional to the vorticity at equilibrium, while the space-time mixing term (corresponding to the boost generator) is proportional to the acceleration. At global equilibrium the four acceleration is the only one consistent with the vorticity profile (providng the necessary centripetal force and allowing a rotation). Indeed, even the temeperature gradient is proportional to the acceleration. The only gradient we need to consider in global rotating equilibrium is then the vorticity, the other ones can be extracted from it. Therefore we include in the Lagrangian density the only gradient necessary for equilibrium.

\section{The dynamics}\label{dynamics}

The most common way to study the evolution of a fluid is to extract from the Lagrangian the stress energy tensor $T^{\mu\nu}$, and close the system of equations using the equation of state and local four-momentum conservation.
From the Lagrangian a generic Lagrangian density an energy-momentum tensor can be constructed~\cite{weinberg}. However, a crucial difference between polarized and unpolarized hydrodynamics is that in the former, due to lack of isotropy, conservation of energy-momentum does not close the equations of motion. 

From the Lagrangian, we can of course match the number of unknowns and equations, but at the price of promoting spin waves to independent degrees of freedom, which will generally violate local entropy maximization.   Since, however, local equilibrium requires spin and vorticity to be aligned,
  Eq. \ref{prevortspin} and the ansatz for $F$ in~\ref{rescaledf} reduce the whole system to three degrees of freedom. The Lagrangian coordinates $\phi^I(x)$. One can than use
the Hamilton principle of action minimization, 
\[ \delta \int d^4 x \mathcal{L}=0 \]
with the proviso that the functional implementation of this principle will lead to a generalization of the usual Euler-Lagrange equations since this Lagrangian in our case depends on second derivatives of fields.  Given there is no explicit dependence on the fields themselves, rather than their derivatives, the correct equation is
 \begin{equation}\label{EL}
 \partial_\mu\partial_\nu\frac{\partial F}{\partial(\partial_\mu\partial_\nu\phi^I)} = \partial_\mu \frac{\partial F}{\partial (\partial_\mu\phi^I)}.
\end{equation}
Since

\begin{equation}
\frac{\partial^2 F}{\partial(\partial_\mu\partial_\nu\phi^I)} = 4\, c\, F^\prime \,\chi  \left( \vphantom{\frac{}{}} \chi + 2 \, \omega^2 \partial_{\omega^2}\chi \right) \omega_{\alpha\beta} \, g^{\alpha\{\mu} P_I^{\nu\}\beta}, \end{equation}

 \[\ \frac{\partial F}{\partial(\partial_\mu\phi^I)} = - F^{\prime} \left[ u_\rho P^{\rho\mu}_I \left(   1-c y^2 - 2  c  b  \chi  \omega^2 \, \partial_b \chi \right)\right]  \]
 \[\ - 2 c \left(   \chi + 2 \, \omega^2 \, \partial _{\Omega^2} \chi \right)F^{\prime} \times \] 

\[\  \times \left\{ \left[ \chi \, \omega^2 -\frac{1}{b}y_{\rho\sigma} \left(   u_\alpha \partial^\alpha K^\rho - u_\alpha \nabla^\rho K^\alpha \right) \right] P^{\sigma\mu}_I - \right.   \] 
 \begin{equation} \left. - \frac{1}{6 b} y_{\rho\sigma}\varepsilon^{\mu\rho\alpha\beta}\epsilon_{IJK}\nabla^\sigma\partial_\alpha \phi^J \partial_\beta \phi^K \right\}.
\end{equation}

This  leads to three conservation law equations, $\partial_\mu J_I^\mu=0$, where
\[\
 J^\mu_I = 4\, c \, \partial_\nu \left\{ F^\prime \left[ \chi  \left(   \chi + 2 \,  \partial _{\Omega^2}\chi \right) \omega_{\alpha\beta} \, g^{\alpha\{\mu} P_I^{\nu\}\beta} \right]\right\} -
\]
 \[\ - F^{\prime} \left[ u_\rho P^{\rho\mu}_I \left(   1-c y^2 - 2  c  b  \chi  \omega^2 \, \partial_b \chi \right)\right] - 2 c \left(   \chi + 2 \, \omega^2 \, \partial _{\Omega^2} \chi \right)F^{\prime} \times \] 

\[\  \times \left\{ \left[ \chi \, \omega^2 -\frac{1}{b}y_{\rho\sigma} \left(   u_\alpha \partial^\alpha K^\rho - u_\alpha \nabla^\rho K^\alpha \right) \right] P^{\sigma\mu}_I - \right.   \] 
 \begin{equation} \left. - \frac{1}{6 b} y_{\rho\sigma}\varepsilon^{\mu\rho\alpha\beta}\epsilon_{IJK}\nabla^\sigma\partial_\alpha \phi^J \partial_\beta \phi^K \right\}.
\end{equation}
with $P^{\mu \nu}_K = \partial K^\mu /\partial(\partial^\nu \phi^K)$, $\nabla^\alpha = \Delta^{\alpha \beta} \partial_\beta$ and $[...],\{...\}$ corresponding to, respectively, antisymmetrization and symmetrization of the indices, as done in \cite{baier}.
  
In addition to generally breaking isotropy and the circulation theorem, unlike non-polarized case the higher gradient of the four velocity will be the third one (fourth one in the fields $\phi^I$).
This system of equation has no easy solutions, however the situations is much simper if one considers the small perturbations from a static background, as it has already be done for the non polarized case (see for instance~\cite{nicolis3}). To understand the consequences of this, we linearize the hydrostatic limit, with a back ground (leading order) entropy density $b_0$
  \begin{equation}
\phi^I = b_0^{1/3} \left[ \delta^I_\mu \, x^\mu +\pi^I(t,{\bf x}) \right]
  \end{equation}
  Using the notation in~\cite{gripaios} we can use as definitions (written in the rest frame of the hydrostatic background) $\dot \pi^I = \partial_t \pi^I=\partial _t \phi^I$, while the contraction $\pi\cdot \partial$ stands for $\delta^\mu_I \pi^I \partial_\mu$ and $[\partial \pi\cdots\partial\pi]$ is a short hand notation for the trace $\delta_J^i\partial_i\pi\cdots\partial \pi^J$. Note that, since the lower-case indices are Lorentz indices, while the upper case (and only latin) ones are internal indices that do not change under a coordinate change, all these definitions become more complicated in other reference frames. It is however convenient in this situation to write everything in this particular frame. We can add another short-hand nontation $\pi\cdot\pi = \sum_I \pi^I \pi^I $. The non-polarized hydrodynamics gives the usual wave equation for
  sound waves, the stationary vortex state  polarization terms which will increase the gradients at each order by one unit.  The free part of the equation (second order in the small fields $\pi^I$) will be, up to an additional  $F(b_0)$ constant which is not relevant for the equations of motion,
  \begin{equation}
 F \simeq  A \left\{ [\partial\pi]  -\frac{1}{2}[\partial\pi\cdot\partial\pi] -\frac{1}{2} \dot\pi^2 \right\} + \end{equation}
\[\ + B \left\{ \vphantom{\frac{}{}} (\partial_\rho \dot\pi)\cdot(\partial^\rho \dot\pi) +  [\partial\dot \pi \cdot \partial \dot \pi]  \right\}
 +\left(\frac{1}{2}A + C\right)[\partial\pi]^2.
\]
and the constants $A,B,C$ are obtained by Taylor-expanding the lagrangian around the usual hydrostatic limit
\begin{equation}
 A = b_0 F^\prime(b_0), \qquad B = A \, c \, \chi^2(b_0,0), \qquad C = \frac{1}{2}b_0^2 F^{\prime\prime}(b_0),
\end{equation}
At the level of the action, the part independent of B is equivalent to that obtained in \cite{nicolis1}, as can be verified by an integration by parts.

If one is interested in the quantum corrections to fluid dynamics, these equations provide the free part of the theory. The lowest order interacting part of the expansion around small perturbation of a static background, that is the third order contribution of the effective Lagrangian of the fields $\pi^I$, is the integral of

\begin{widetext}
\begin{eqnarray}
 L_3 &\simeq&  A \left\{  \frac{1}{6}[\partial\pi\cdot\partial\pi\cdot\partial\pi] -\frac{1}{4}[\partial\pi][\partial\pi\cdot\partial\pi] \right.  \nonumber \\
 && \qquad   + (\dot\pi\cdot\partial\pi)\cdot\pi - \frac{1}{2}[\partial\pi]\dot\pi^2 + c \,\chi^2(b_0,0) \left[ \vphantom{\frac{}{}} (\partial_\mu \dot \pi)\cdot(\partial^\mu \dot \pi) + [\partial\dot \pi \cdot \partial \dot \pi] -(\partial_\mu \dot \pi)\cdot( \partial^\mu \dot \pi \cdot \partial\pi)\right. \nonumber \\
 && \quad \left. \vphantom{\frac{}{}}  -(\partial_\mu \dot \pi)\cdot \{\dot\pi\cdot\partial (\partial^\mu \pi)\} -  2 (\ddot\pi\cdot\partial \dot \pi)\cdot \dot \pi - [\partial\dot \pi \cdot \partial \dot \pi\cdot \partial \pi] - \dot \pi \cdot \partial(\partial_I \pi^J)\partial_J \dot \pi^I + (\dot \pi \cdot \partial \dot \pi)\cdot  \ddot \pi  +  \ddot \pi \cdot \dot\pi \cdot \ddot \pi \right] \nonumber \\
 && \qquad  \left. \vphantom{\frac{1}{2}} + c \,\chi(b_0,0)\left[ \vphantom{\frac{}{}}\chi(b_0,0) + 2 \, b_0 \, \partial_b\chi(b_0, 0) \right] \left[ \vphantom{\frac{}{}}  [\partial\pi](\partial_\mu \dot \pi)\cdot(\partial^\mu \dot \pi) + [\partial\pi][\partial\dot \pi \cdot \partial \dot \pi]  \right] \right\} \nonumber \\ \nonumber \\
&& + C [\partial\pi] \left\{ \vphantom{\frac{1}{2}} [\partial\pi]^2 - [\partial\pi\cdot\partial\pi] - \dot\pi^2 + 2 \, c \, \chi^2(b_0,0)\left[ \vphantom{\frac{}{}} (\partial_\mu\dot \pi)\cdot(\partial^\mu\dot\pi) + [\partial\dot\pi\partial\dot\pi] \right] \right\} \nonumber \\ \nonumber \\
&& +\frac{1}{6} b_0^3 F^{\prime\prime\prime}(b_0)\, [\partial\pi]^3.
\end{eqnarray}
\end{widetext}
We note that the Lagrangian becomes, already at leading order with no dissipative corrections, dependent on second derivative terms.   As has been known since the 19th century (Ostrogradski's theorem \cite{ostro}), such lagrangians are inherently unstable, something which can be used, in the context of dissipative hydrodynamics, for motivating the introduction of non-hydrodynamic degrees of freedom \cite{torrieri3}.  The presence of higher order gradient terms at the ''lowest level'', therefore, means that dissipative corrections or the appearance of new degrees of freedom become necessary to preserve the hydrostatic vaccuum even in the ideal limit, a realization that we explore in detail in \cite{prl}.

What this means is that the instabilities plaguing such a higher-order system could lead to a local ``thermalization'' between hydrodynamic and polarizing degrees of freedom, imposing an effective viscosity also on ``ideal'' fluid dynamics systems.   This idea, related to the existance of a lower limit of viscosity \cite{kss}, will be explored in a subsequent paper \cite{prl}.
\begin{figure*}
\epsfig{width=0.6\textwidth,figure=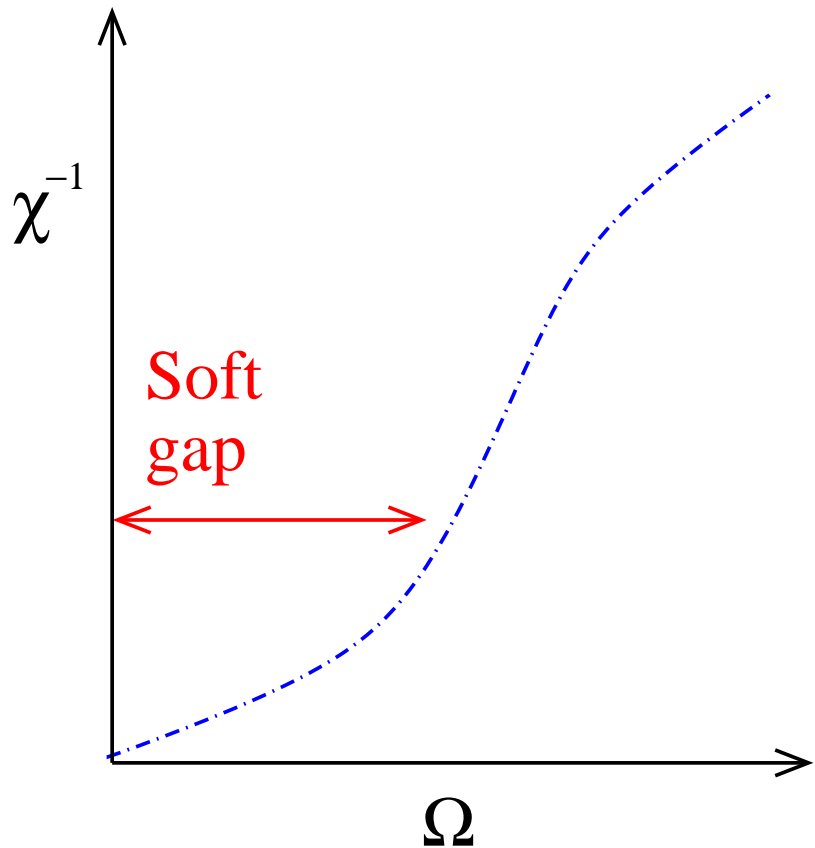}
\caption{\label{vort}A thermodynamically sensible relationship betwee $\chi$ and vorticity}
\end{figure*}
On the other had the non-linear terms, however, all depend on derivatives of $\chi(b,\omega^2)$ which are expected, for sensitive equations of state, to be high at small vorticity and diminish at high vorticity, when particle polarizations saturates and cannot contribute anylonger to the angular momentum of the fluid cell (Fig. \ref{vort}. At small vorticity creating polarization is preferable than creating vortices, at large vorticity this effect is small).  Hence, we expect this non-linearity to create an effective ``soft energy gap'' for vortices, ensuring they only form when applied angular momentum is large enough.  Such a gap could alleviate the instabilities seen in \cite{nicolis1} and  
it will be interesting to see if a more quantitative estimate of this effect can be made.

In conclusion, we developed the effective theory for hydrodynamics in the limit where the mean free path is negligible but the microscopic degrees of freedom exhibit microscopic polarization.   This theory is likely to be highly relevant to the phenomenology of global polarization of hadronic collisions \cite{upsal} and might have an impact to the description of chiral observables \cite{magnetic1}.  The third order gradient nature of this theory might also impact the question of weather a lower quantum viscosity limit is realized in nature.   We hope understanding of all these areas will increase in the coming years.

\section*{Acknowledgements}
\textit{Acknowledgements} GT acknowledges support from FAPESP proc. 2014/13120-7 and CNPQ bolsa de 
produtividade 301996/2014-8. LT was 
supported by  the  U.S.  Department  of  Energy,  Office  of  
Science,  Office of Nuclear Physics under Award No. DE-SC0004286   
and  Polish National Science Center Grant DEC-2012/06/A/ST2/00390. DM would like to acknowledge CNPQ graduate fellowship n. 147435/2014-5
Parts of this work were done when LT visited Campinas on FAEPEX fellowship number 2020/16, as well as when GT participated in the INT workshop "Exploring the QCD Phase Diagram through Energy Scans
"  We thank FAEPEX and the INT organizers for the support provided.
We wish to thank Miklos Gyulassy for enlightening discussions which posed the conceptual challenges that eventually led to this work,  and Mike Lisa for showing us experimental literature and useful discussions.


\end{document}